\documentclass[letterpaper,11pt]{article}
 \usepackage[utf8]{inputenc}
\usepackage{caption} 
\usepackage[margin=2.5cm]{geometry}
\usepackage{amsmath,amssymb,amsfonts}
\usepackage{algorithmic}
\usepackage{graphicx}
\usepackage{textcomp}
\usepackage{listings}
\usepackage{tikz}

\usepackage{forest}
\usepackage{tikz-qtree}
\usepackage[linesnumbered,lined,boxed,commentsnumbered]{algorithm2e}
\usepackage{balance}
\usepackage{multirow}

\usepackage{caption}
\DeclareCaptionFont{ieeeblue}{\color{accessblue}}
\newcommand{\figcapfont}[1]{#1}
\DeclareCaptionLabelFormat{myformat}{\figcapfont{\textbf{#1} \textbf{#2}}}
\usepackage{subcaption}

\usepackage{multirow}        
\usepackage{array}           
\usepackage{xcolor}          

\usepackage{multicol}
\usepackage{enumitem}
\usepackage{url}
\usepackage{etoolbox}
\usepackage{stmaryrd}
\usepackage{fancyvrb}
\definecolor{CypherGreen}{RGB}{0, 120, 0}
\definecolor{bblue}{HTML}{4F81BD}
\definecolor{rred}{HTML}{C0504D}
\definecolor{ggreen}{HTML}{9BBB59}
\definecolor{ppurple}{HTML}{9F4C7C}
\definecolor{accessblue}{HTML}{4F81BD}
\definecolor{greycolor}{cmyk}{0,0,0,.8}

\usepackage{cprotect}

\usepackage{pgfplots}
\usepackage{pgfplotstable}
\pgfplotsset{compat=1.17}

\usepackage{comment}
\usepackage{tabularx}
\usepackage{booktabs}

\newcommand{\code}[1]{{\texttt{\footnotesize#1}}}

\newcommand{\setO}{\textnormal{\textbf{O}}\xspace}
\newcommand{\setL}{\textnormal{\textbf{L}}\xspace}
\newcommand{\setV}{\textnormal{\textbf{V}}\xspace}
\newcommand{\setN}{\mathcal{N}}
\newcommand{\setE}{\mathcal{E}}
\newcommand{\node}{\mathit{Node}}
\newcommand{\edge}{\mathit{Edge}}

\newcommand{\sequence}{\mathit{Seq}}
\newcommand{\len}{\mathit{Len}}
\newcommand{\lab}{\mathit{Label}}

\newcommand{\dom}{\mathit{Dom}}
\newcommand{\paths}{\mathit{Paths}}
\newcommand{\true}{\mathtt{true}} 
\newcommand{\false}{\mathtt{false}}
\newcommand{\nulo}{\mathtt{null}}

\newcommand{\tm}{\ensuremath{\mathcal{T}_{M}}}
\newcommand{\tw}{\ensuremath{\mathcal{T}_{W}}}
\newcommand{\tr}{\ensuremath{\mathcal{T}_{R}}}

\newcommand{\tq}{\ensuremath{\mathcal{T}_{Q}}}

\newcommand{\eval}{\ensuremath{\mathit{Eval}}}

\newcommand{\proy}[2]{\ensuremath{proy(#1 , #2)}}

\def\BibTeX{{\rm B\kern-.05em{\sc i\kern-.025em b}\kern-.08em
    T\kern-.1667em\lower.7ex\hbox{E}\kern-.125emX}}

\newcolumntype{R}{>{\raggedleft\arraybackslash}r}
\definecolor{clrWalk}   {RGB}{  0, 84,166}   
\definecolor{clrTrail}  {RGB}{227,114,  0}   
\definecolor{clrSimple} {RGB}{ 39,139, 34}   
\definecolor{clrAcyclic}{RGB}{163,  0, 44}   

\begin{document}

\title{PathDB: A System for Evaluating Regular Path Queries}

\author{
    Roberto García$^{1,2}$, 
    Renzo Angles$^{1,2}$, 
    Vicente Rojas$^{2}$, 
    Sebastián Ferrada$^{2,3,4}$ \\[2ex]
    \small $^{1}$Department of Computer Science, Faculty of Engineering, Universidad de Talca, Curic\'o, Chile \\
    \small $^{2}$Millennium Institute for Foundational Research on Data (IMFD), Santiago, Chile \\
    \small $^{3}$Data and Artificial Intelligence Initiative, Universidad de Chile, Santiago, Chile \\
    \small $^{4}$National Center for Artificial Intelligence Research (CENIA), Santiago, Chile 
}

\date{} 
\maketitle
\begin{abstract}
Regular Path Queries (RPQs) are a core mechanism for expressing recursion and reachability in graph databases. However, most systems evaluate RPQs with traversal-based algorithms that repeatedly explore overlapping subpaths and offer limited control over path semantics. We present PathDB, an algebraic query engine for RPQs based on a closed path algebra over multisets of paths with five operators: selection, join, union, recursive join, and projection. PathDB provides (i) a GQL-inspired declarative language that supports RPQs with multiple semantics (walk, trail, simple, and acyclic), (ii) an operator-at-a-time execution procedure analogous to relational query processing, and (iii) result formats that include full paths, not only endpoint pairs. 
The experimental evaluation, based on four LDBC Social Network Benchmark property graphs and a workload of 142 queries derived from 26 path patterns, showed that PathDB outperforms two automaton-guided traversal baselines (DFS and BFS), often by more than an order of magnitude.
\end{abstract}


\section{Introduction}
Graph databases are commonly used to manage highly connected data in applications such as social networks, knowledge graphs, and biological networks\cite{sakr2021future}. Compared with relational databases, graph database systems model entities and relationships explicitly, enabling efficient relationship-centric queries. As a result, graph query languages include path queries that retrieve sequences of edges (and their incident nodes) that match structural constraints.

Regular Path Queries (RPQs) are a powerful mechanism for querying paths in graph databases. RPQs let users describe path patterns as regular expressions over edge labels and retrieve paths whose label sequences match the expression~\cite{Bonifati2018}. This capability is essential for expressing recursive patterns such as reachability and transitive relationships~\cite{AnglesABHRV17}. Consequently, RPQs and their extensions have become central features of practical graph query languages, including SPARQL~\cite{sparql2013}, Cypher~\cite{francis2018}, GQL~\cite{gql2024}, and SQL/PGQ~\cite{sqlpgq2023}.

Despite their expressive power, evaluating RPQs efficiently remains challenging. Most graph database systems evaluate RPQs with traversal-based algorithms, typically breadth-first search (BFS) or depth-first search (DFS) variants guided by a finite automaton compiled from the regular expression. While conceptually simple, this approach often redundantly explores shared subpaths, especially in graphs with high branching or many cycles. As recursion deepens, traversal-based algorithms may revisit the same intermediate states many times, leading to substantial overhead~\cite{farias2023}.

An alternative is to evaluate path queries with algebraic operators, analogous to relational algebra. In an algebraic framework, path queries can be expressed as compositions of operators that manipulate sets (or multisets) of paths. This perspective treats evaluation as an operator tree and enables classical optimizations such as predicate pushdown, operator reordering, and cost-based planning. However, few graph database systems provide a fully algebraic model for constructing and evaluating paths.

In this paper, we present PathDB, a query engine that evaluates RPQs using a closed algebra over multisets of paths. Rather than relying on traversal-based evaluation, PathDB executes queries as compositions of algebraic operators that construct, combine, and filter candidate paths. The algebra captures recursion through a recursive join operator, which systematically constructs result paths while supporting configurable semantics (walk, trail, simple, and acyclic).

PathDB is implemented in Java and includes a command-line interface (CLI), a storage manager, and a query processor. The CLI lets users submit path queries in a declarative language inspired by the emerging GQL standard~\cite{gql2024}. The storage manager keeps the property graph in memory using hash-table-based structures. The query processor compiles declarative path queries into logical plans based on the proposed path algebra, then translates them into iterator-based physical plans. This architecture represents path queries as operator trees, which simplifies extensibility and supports future optimizations.

To assess the approach, we conducted an experimental study on a large-scale dataset generated by the LDBC Social Network Benchmark. We compare PathDB against two automaton-guided traversal baselines. The results show that PathDB performs consistently better across a wide range of RPQ patterns, especially for recursive queries where traversal-based methods incur substantial redundancy.

This paper makes three main contributions:
\begin{enumerate}
\item \textit{A closed path algebra for RPQs}.
We introduce an algebraic framework for querying paths in property graphs, with operators for selection, join (concatenation), union, projection, and recursive join.
\item \textit{An algebra-based evaluation strategy for RPQs}.
We show how to compile RPQs into algebraic expressions and execute them using operator-based plans rather than traversal.
\item \textit{An experimental evaluation on large-scale graph data}.
We evaluate PathDB on the LDBC Social Network Benchmark and compare it against traversal-based baselines.
\end{enumerate}

The rest of the paper is organized as follows. Section~\ref{sec:sysOver} overviews PathDB, including the storage manager, query processor, and client manager. Section~\ref{sec:graphstorage} describes the data model and the in-memory graph representation. Section~\ref{sec:algebra} presents the path algebra used to evaluate RPQs. Section~\ref{sec:language} introduces the declarative query language. Section~\ref{sec:expeval} reports the experimental evaluation. Section~\ref{sec:relatedwork} discusses related work. Finally, Section~\ref{sec:conclusion} concludes and describes future work.

\section{System Overview}
\label{sec:sysOver}
\begin{figure}[htbp]
    \centering
    \includegraphics[width=0.5\textwidth]{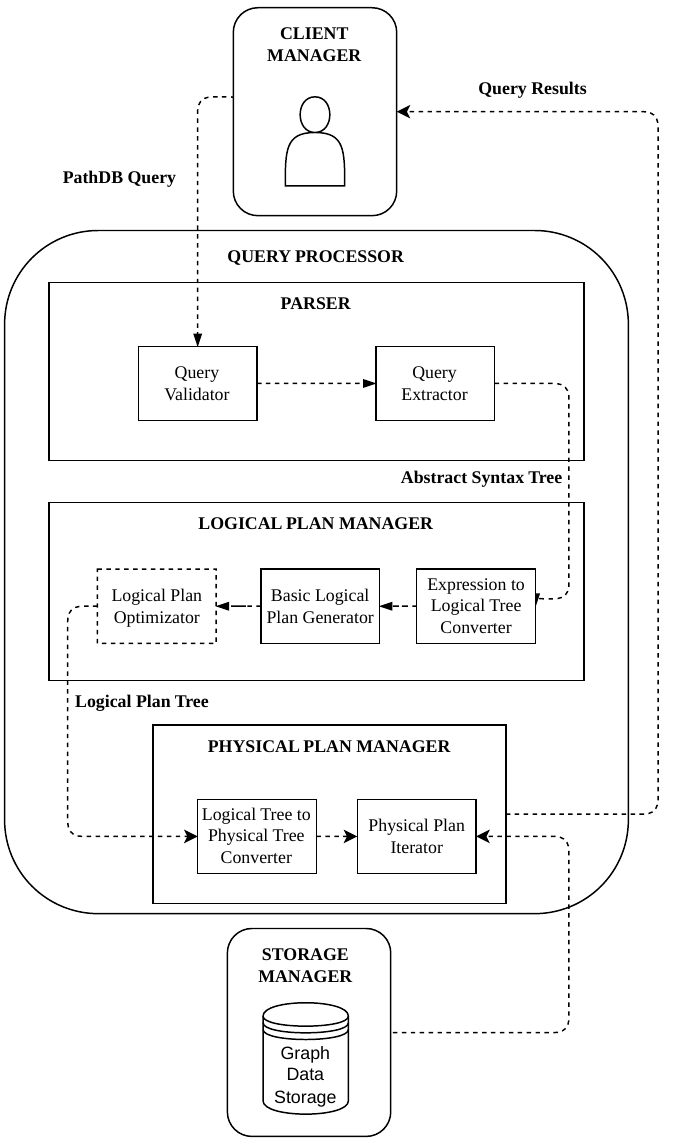}
    \caption{PathDB architecture.}
    \label{fig:systemArch}
\end{figure}

PathDB follows a modular architecture inspired by the classical structure of database management systems~\cite{Hellerstein2007}. As shown in Figure~\ref{fig:systemArch}, PathDB comprises three main modules: the Storage Manager, the Query Processor, and the Client Manager.

PathDB is implemented in Java to support a modular execution engine based on standard database design patterns, such as iterator-based operators and visitor-style plan traversal. Java’s managed runtime and object-oriented abstractions make these patterns straightforward to implement while preserving a clear separation of responsibilities.

\subsection{Storage manager}
The Storage Manager represents a property graph entirely in main memory, following the model described in Section~\ref{sec:graphstorage}. The current implementation uses hash-table-based structures within the Java Virtual Machine heap, storing nodes and edges as interconnected objects. Nodes are indexed by identifier to provide constant-time access, while edges are grouped by label to support efficient retrieval during path evaluation. This deliberately minimal design matches PathDB’s path-centric processing model, where edge-label lookups dominate execution, and enables efficient navigation through direct in-memory references.

\subsection{Query processor}
The Query Processor transforms user queries into executable plans and produces results. It comprises three components: the parser, the logical plan manager, and the physical plan manager.

The parser takes a query expressed in PathDB’s GQL-inspired language and produces an abstract syntax tree (AST). The logical plan manager then transforms the AST into a logical plan based on our recursive path algebra. This transformation follows a direct correspondence between RPQ syntax and algebraic operators, as described in Section~\ref{sec:semantics}. The resulting logical plan captures the declarative semantics of the query and serves as input for physical planning.

The physical plan manager translates the logical plan into a physical plan represented as a tree of executable operators. The current implementation supports basic physical operators, including sequential scans and hash joins, and maps each logical operator to a corresponding physical one (cf. Section~\ref{sec:language}). More advanced algorithms, as well as a dedicated physical planner and optimizer, are part of ongoing work.

The physical plan manager executes the physical plan using a lazy, iterator-based model. Each operator implements a common interface exposing \code{hasNext}, \code{next}, and \code{close} methods, enabling pipelined execution and on-demand result production. This design also naturally supports language features such as \code{LIMIT}. The Java implementation facilitates this approach through polymorphism, visitor patterns, and clear abstraction boundaries between operators.

Overall, PathDB follows a layered execution pipeline in which the parser, logical plan manager, physical plan manager, and storage manager interact through well-defined interfaces. Plans are executed via iterator-based access to the underlying graph, promoting modularity and making it straightforward to extend the system with new operators or execution strategies.

\subsection{Client manager}
The Client Manager provides the user-facing interface to PathDB and is currently realized as a command-line interface (CLI). PathDB is distributed as a standalone JAR file, allowing users to load property graphs into main memory and execute path queries interactively. Query results are displayed in tabular form, together with basic execution statistics such as the number of solution paths and query runtime.

In addition to a default example graph, the Client Manager supports loading custom property graphs specified in the Property Graph Data Format (PGDF)~\cite{10734106}. This interface decouples user interaction from query processing and storage concerns, completing PathDB’s modular architecture.

\section{Graph Storage}
\label{sec:graphstorage}
In this section, we describe PathDB’s graph storage module, including the underlying data model and the in-memory data structures used to store property graphs.

\subsection{Data Model}
\label{sec:data_model}
\begin{figure} []
  \centering
    \includegraphics[width=0.7\linewidth]{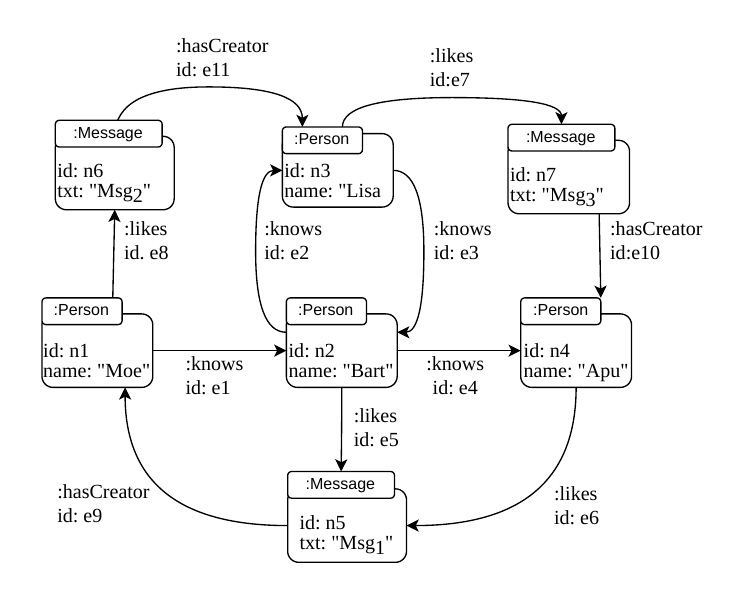}
  \caption{Graphical representation of a property graph modeling a social network.}
  \label{fig:graph}
\end{figure}

PathDB targets property graphs, i.e., labeled directed graphs whose nodes and edges may carry properties. Figure~\ref{fig:graph} illustrates an example of property graph. Next, we formalize the data model.

Let \setO be an infinite set of object identifiers, \setL an infinite set of labels, and \setV an infinite set of values (numbers, strings, date-times, etc.). We represent labels in \setL as unquoted strings without whitespace, and values in \setV as quoted strings.

Formally, a \emph{property graph} is a tuple $G = (\setN, \setE, \rho, \lambda, \nu)$ where:
\begin{enumerate}
\item 
$\setN \subset$ \setO is a finite set of node identifiers;
\item 
$\setE \subset$ \setO is a finite set of edge identifiers, disjoint with $\setN$, i.e., $\setN \cap \setE = \emptyset$;
\item 
$\rho : \setE \xrightarrow[]{} (\setN \times \setN)$ is a total function that assigns a pair of nodes to each edge;
\item 
$\lambda : (\setN \cup \setE) \xrightarrow[]{}$ \setL is a total function that assigns a label to each node or edge;
\item $\nu: (\setN \cup \setE)\ \times$  \setL $\rightharpoonup$ \setV is a partial function that defines properties for nodes and edges.
\end{enumerate}

Given an edge $e \in \setE$, if $\rho(e) = (n_1, n_2)$ then $n_1$ is the \emph{source} of $e$ and $n_2$ is its \emph{target}. If $\nu(o,l) = v$, then the object $o$ (node or edge) has property $l$ with value $v$.

Given a property graph $G = (\setN, \setE, \rho, \lambda, \nu)$, a \emph{path} $p$ in $G$ is a sequence $[n_1, e_1, n_2, e_2, \dots, e_k, n_{k+1}]$ where $k \ge 0$, $n_i \in \setN$, and $e_i \in \setE$, with $\rho(e_i) = (n_i, n_{i+1})$. The functions $\node(p,i)$ and $\edge(p,i)$ return the $i$-th node (resp. edge) identifier in $p$, and $\sequence(p)$ returns the sequence of identifiers occurring in $p$. The length of $p$, denoted $\len(p)$, is the number of edges in $p$. The label of $p$ is the concatenation of the edge labels along $p$, i.e., $\lab(p) = \lambda(e_1) \circ \dots \circ \lambda(e_k)$; if $p$ has no edges, then $\lab(p)$ is the empty string.

\subsection{Graph Data Structure}
\label{sec:graphdatastructure}
The current version of PathDB evaluates path queries over property graphs stored in main memory (RAM). Since PathDB is implemented in Java, a property graph is represented as a collection of interconnected objects on the Java Virtual Machine (JVM) heap. This design leverages direct in-memory references to support efficient graph access during query processing.

The object model consists of three primary classes: \emph{Graph}, \emph{Node}, and \emph{Edge}. Each object carries a map of attributes to satisfy the property graph model defined in Section~\ref{sec:data_model}. The \emph{Node} class stores a unique identifier, a label, and a set of properties. The \emph{Edge} class stores a unique identifier, a label, its source and target nodes, and a set of properties (implemented as a hash map of pairs $\langle$property name, property value$\rangle$).

The \emph{Graph} class maintains collections of nodes and edges. Nodes are stored in a hash-based index that maps node identifiers to references to the corresponding node objects. Figure~\ref{fig:node_storage} illustrates this index for a subset of the graph in Figure~\ref{fig:graph}. Edges are stored in a separate hash table that maps each edge label to a linked list of edges with that label (Figure~\ref{fig:edge_storage}).

Note that hash tables enable expected constant-time node lookup~\cite{10.1145/3519935.3519969}. Moreover, the edge-label index matches PathDB’s access pattern, since evaluating path algebra expressions primarily requires retrieving edges by label.

\begin{figure}[t!]
  \centering
\begin{tikzpicture}[
  keycell/.style={draw, rectangle, minimum width=1.5cm, minimum height=0.6cm, font=\small},
  valcell/.style={draw, rectangle, minimum width=1.5cm, minimum height=0.6cm},
  value/.style={circle, fill=black, inner sep=2pt},
  nodebox/.style={draw, rectangle, minimum width=3.5cm, minimum height=0.6cm, font=\scriptsize},
  arrow/.style={->, thick}
  ]

  \node[keycell, font=\bfseries] (keyh) at (0,2.0) {Key};
  \node[valcell, font=\bfseries] (valh) at (1.5,2.0) {Value};

  \node[keycell] (k1) at (0,1.5) {n1};
  \node[valcell] (vc1) at (1.5,1.5) {};
  \node[value] (v1) at (1.5,1.5) {};

  \node[keycell] (k2) at (0,1.0) {n2};
  \node[valcell] (vc2) at (1.5,1.0) {};
  \node[value] (v2) at (1.5,1.0) {};

  \node[keycell] (k3) at (0,0.5) {n3};
  \node[valcell] (vc3) at (1.5,0.5) {};
  \node[value] (v3) at (1.5,0.5) {};

  \node[keycell] (k4) at (0,0.0) {n4};
  \node[valcell] (vc4) at (1.5,0.0) {};
  \node[value] (v4) at (1.5,0.0) {};

  \node[nodebox] (n1) at (6,2.0) {label:Person, props:\{name:"Moe"\}};
  \node[nodebox] (n2) at (6,1.0) {label:Person, props:\{name:"Bart"\}};
  \node[nodebox] (n3) at (6,0.0) {label:Person, props:\{name:"Lisa"\}};
  \node[nodebox] (n4) at (6,-1.0) {label:Person, props:\{name:"Apu"\}};

  \draw[arrow, dashed] (v1.east) to[out=0,in=180,looseness=1.2] (n1.west);
  \draw[arrow, dashed] (v2.east) to[out=0,in=180,looseness=1.2] (n2.west);
  \draw[arrow, dashed] (v3.east) to[out=0,in=180,looseness=1.2] (n3.west);
  \draw[arrow, dashed] (v4.east) to[out=0,in=180,looseness=1.2] (n4.west);

\end{tikzpicture}
\caption{Hash-based index for node storage. Each entry contains a node identifier as key, and the value is a reference to the corresponding node object.}
\label{fig:node_storage}
\end{figure}

\begin{figure}[t!]
  \centering
\begin{tikzpicture}[
  keycell/.style={draw, rectangle, minimum width=2cm, minimum height=0.6cm, font=\small},
  valcell/.style={draw, rectangle, minimum width=1.3cm, minimum height=0.6cm},
  value/.style={circle, fill=black, inner sep=2pt},
  edgebox/.style={draw, rectangle, minimum width=1.2cm, minimum height=0.6cm},
  objbox/.style={draw, rectangle, minimum width=6cm, minimum height=0.6cm, font=\small},
  arrow/.style={->, thick}
  ]

  \node[keycell, font=\bfseries] (keyh) at (0,1.5) {Key};
  \node[valcell, font=\bfseries] (valh) at (1.65,1.5) {Value};

  \node[keycell] (k1) at (0,1.0) {knows};
  \node[valcell] (vc1) at (1.65,1.0) {};
  \node[value] (v1) at (1.65,1.0) {};

  \node[keycell] (k2) at (0,0.5) {likes};
  \node[valcell] (vc2) at (1.65,0.5) {};
  \node[value] (v2) at (1.65,0.5) {};

  \node[keycell] (k3) at (0,0.0) {hasCreator};
  \node[valcell] (vc3) at (1.65,0.0) {};
  \node[value] (v3) at (1.65,0.0) {};

  \node[edgebox] (a1) at (4,1.5) {};
  \draw ([xshift=-0.3cm]a1.north east) -- ([xshift=-0.3cm]a1.south east);
  \node[circle, fill=black, inner sep=1.5pt] (a1dot) at ($(a1.west)!0.5!(a1.center)$) {};

  \node[edgebox] (a2) at (5.5,1.5) {};
  \draw ([xshift=-0.3cm]a2.north east) -- ([xshift=-0.3cm]a2.south east);
  \node[circle, fill=black, inner sep=1.5pt] at ($(a2.west)!0.5!(a2.center)$) {};

  \node[edgebox] (a3) at (7,1.5) {};
  \draw ([xshift=-0.3cm]a3.north east) -- ([xshift=-0.3cm]a3.south east);
  \node[font=\scriptsize] at ($(a3.west)!0.5!(a3.center)$) {null};

  \node[edgebox] (c1) at (4,0.5) {};
  \draw ([xshift=-0.3cm]c1.north east) -- ([xshift=-0.3cm]c1.south east);
  \node[circle, fill=black, inner sep=1.5pt] at ($(c1.west)!0.5!(c1.center)$) {};

  \node[edgebox] (c2) at (5.5,0.5) {};
  \draw ([xshift=-0.3cm]c2.north east) -- ([xshift=-0.3cm]c2.south east);
  \node[circle, fill=black, inner sep=1.5pt] at ($(c2.west)!0.5!(c2.center)$) {};

  \node[edgebox] (c3) at (7,0.5) {};
  \draw ([xshift=-0.3cm]c3.north east) -- ([xshift=-0.3cm]c3.south east);
  \node[font=\scriptsize] at ($(c3.west)!0.5!(c3.center)$) {null};

  \node[edgebox] (b1) at (4,-0.5) {};
  \draw ([xshift=-0.3cm]b1.north east) -- ([xshift=-0.3cm]b1.south east);
  \node[circle, fill=black, inner sep=1.5pt] at ($(b1.west)!0.5!(b1.center)$) {};

  \node[edgebox] (b2) at (5.5,-0.5) {};
  \draw ([xshift=-0.3cm]b2.north east) -- ([xshift=-0.3cm]b2.south east);
  \node[circle, fill=black, inner sep=1.5pt] at ($(b2.west)!0.5!(b2.center)$) {};

  \node[edgebox] (b3) at (7,-0.5) {};
  \draw ([xshift=-0.3cm]b3.north east) -- ([xshift=-0.3cm]b3.south east);
  \node[font=\scriptsize] at ($(b3.west)!0.5!(b3.center)$) {null};

  \node[objbox] (obj1) at (3.5,2.6) {id:e1, label:knows, src:n1, tgt:n2};

  \draw[arrow, dashed] (v1.east) to[out=0,in=180,looseness=1.2] (a1.west);
  \draw[arrow, dashed] (v2.east) to[out=0,in=180,looseness=1.2] (c1.west);
  \draw[arrow, dashed] (v3.east) to[out=0,in=180,looseness=1.2] (b1.west);

  \draw[arrow] (a1.east) -- (a2.west);
  \draw[arrow, dotted] (a2.east) -- (a3.west);

  \draw[arrow] (c1.east) -- (c2.west);
  \draw[arrow, dotted] (c2.east) -- (c3.west);

  \draw[arrow] (b1.east) -- (b2.west);
  \draw[arrow, dotted] (b2.east) -- (b3.west);

  \draw[arrow] (a1dot.north) to[out=90,in=270] (obj1.south);

\end{tikzpicture}

\caption{Hash-based index for edge storage. Each entry contains an edge label as key, and the value is a reference to a linked list of edges having that label. Each element of the linked list contains either a reference to an edge object or \emph{null} to indicate the end of the linked list.}
\label{fig:edge_storage}

\end{figure}

\section{Path Algebra}
\label{sec:algebra}
This section presents a path algebra that provides a formal framework for querying and manipulating paths. The algebra underpins PathDB’s evaluation of RPQs by expressing queries as compositions of algebraic operators.

\subsection{Basic definitions}

\subsubsection{Basic path operations}
Let $p_1$ and $p_2$ be paths in a property graph $G = (\setN, \setE, \rho, \lambda, \nu)$.
We say that $p_1$ and $p_2$ are \emph{linkable}, written $p_1 \risingdotseq p_2$, if and only if the last node of $p_1$ equals the first node of $p_2$, i.e., $\node(p_1, \len(p_1)+1) = \node(p_2, 1)$.
If $p_1 = [n_1, e_1, \ldots, n_j]$ and $p_2 = [n_j, e_j, \ldots, n_k]$, their concatenation, denoted $p_1 \circ p_2$, is the path $[n_1, e_1, \ldots, n_j, e_j, \ldots, n_k]$.
For instance, if $p_1 = [n_1, e_1, n_2]$ and $p_2 = [n_2, e_3, n_3]$, then $p_1 \circ p_2 = [n_1, e_1, n_2, e_3, n_3]$.

Two paths $p_1$ and $p_2$ are equal if $\len(p_1)=\len(p_2)$ and their node and edge sequences coincide, i.e., $\node(p_1,i)=\node(p_2,i)$ for $i \in \{1, \ldots, \len(p_1)+1\}$ and $\edge(p_1,j)=\edge(p_2,j)$ for $j \in \{1, \ldots, \len(p_1)\}$.

Additionally, for multisets of paths $S_1$ and $S_2$, we write $S_1 \subseteq S_2$ if every path occurs in $S_1$ with multiplicity at most its multiplicity in $S_2$. We write $S_1 = S_2$ if and only if $S_1 \subseteq S_2$ and $S_2 \subseteq S_1$.

\subsubsection{Path restrictors}
Breadth-First Search (BFS) and Depth-First Search (DFS) are the baseline algorithms for obtaining all the paths in a graph. However, these algorithms have the problem of finiteness when the graph has cycles, i.e., the number of solutions may be infinite, and the algorithm therefore does not terminate. 

To address the finiteness problem, a graph query language can be designed to support specific path semantics.
Path semantics defines the rules governing graph traversal and the criteria that determine which paths are valid for a path query.

In GQL~\cite{gql2024}, \emph{path restrictors} are syntactic keywords that specify the path semantics of a query.
Following GQL, we consider four path restrictors: \textit{Walk}, \textit{Trail}, \textit{Acyclic}, and \textit{Simple}.

Given a path $p$ and a path restrictor $\tau$, we will use $p \models \tau$ to denote that $p$ satisfies $\tau$, and $p \nvDash \tau$ otherwise.
The semantics of $p \models \tau$ is defined as follows:
\begin{itemize}
    \item if $\tau = Walk$, then $p \models \tau$ for every path $p$, without any restriction;
    \item if $\tau = Trail$, then $p \models \tau$ if and only if $p$ contains no repeated edges;
    \item if $\tau = Acyclic$, then $p \models \tau$ if and only if $p$ contains no repeated nodes;
    \item if $\tau = Simple$, then $p \models \tau$ if and only if $p$ contains no nodes, except possibly for the source and target nodes.
\end{itemize}

\subsubsection{Selection Conditions}
Let $i$ be a positive integer, $v \in \setV$ be a value, and $l \in \setL$ be a label. 
A \emph{simple selection condition} is any of the following expressions:\footnote{Our definition of simple selection conditions can be easily extended to support inequalities ($\neq$, $<$, $>$, $\leq$, $\geq$) and built-in functions.}
$label(node(i)) = l$, 
$label(edge(i)) = l$, 
$prop(node(i),l) = v$, 
$prop(edge(i),l) = v$, 
$length() = i$, $isTrail()$, $isSimple()$, and $isAcyclic()$.
If $c_1$ and $c_2$ are selection conditions, then $(c_1 \land c_2)$, $(c_1 \lor c_2)$, and $( \neg c_1 )$ are complex selection conditions.

Let $p$ be a path that occurs in a graph $G = (\setN, \setE, \rho, \lambda, \nu)$.
The evaluation of a selection condition $c$ over $p$, denoted $\Psi(c,p)$, returns $\true$ or $\false$.  A simple condition $c$ is evaluated as $\true$ in the following cases:
\begin{itemize}
\item If $c$ is $label(node(i)) = l$, $i \leq \len(p) + 1$, and $\lambda(\node(p,i)) = l$;
\item If $c$ is $label(edge(i)) = l$, $i \leq \len(p)$, and $\lambda(\edge(p,i)) = l$;
\item If $c$ is $prop(node(i),l) = v$, $i \leq \len(p) + 1$, $(\node(p,i),l) \in \dom(\nu)$, and $\nu(\node(p,i), l) = v$;
\item If $c$ is $prop(edge(i),l) = v$, $i \leq \len(p)$, $(\edge(p,i),l) \in \dom(\nu)$, and $\nu(\edge(p,i), l) = v$;
\item If $c$ is $length() = i$ and $\len(p) = i$;
\item If $c$ is $isTrail()$ and $p \models trail$;
\item If $c$ is $isSimple()$ and $p \models Simple$;
\item If $c$ is $isAcyclic()$ and $p \models Acyclic$.
\end{itemize}
In any other case, $c$ is evaluated as $\false$.  
The evaluation of a complex selection condition follows the usual semantics of propositional logic regarding negation, conjunction, and disjunction.

\subsubsection{Projection Terms}
Let $i$ be a positive integer, and $l \in \setL$ be a label.  
A \emph{projection term} is any of the following expressions: 
$seq()$,
$node(i)$, 
$edge(i)$, 
$nLabel(i)$,  
$eLabel(i)$,  
$nProp(i,l)$,
$eProp(i,l)$,
and
$length()$. 

Let $p$ be a path that occurs in a graph $G = (\setN, \setE, \rho, \lambda, \nu)$.
The evaluation of a projection term $t$ over $p$, denoted $\Theta(t,p)$, is defined as follows:
\begin{itemize}
\item If $t$ is $seq()$ then $\Theta(t,p) = \sequence(p)$;
\item If $t$ is $node(i)$ then $\Theta(t,p) = \node(p,i)$ when $i \leq \len(p) + 1$;
\item If $t$ is $edge(i)$ then $\Theta(t,p) = \edge(p,i)$ when $i \leq \len(p)$ ;
\item If $t$ is $label(node(i))$ then $\Theta(t,p) = \lambda(\node(p,i))$ when $i \leq \len(p) + 1$;
\item If $t$ is $label(edge(i))$ then $\Theta(t,p) = \lambda(\edge(p,i))$ when $i \leq \len(p)$;
\item If $t$ is $prop(node(i),l)$ then $\Theta(t,p) = \nu(\node(p,i), l)$ when $i \leq \len(p) + 1$, and 
$(\node(i),l) \in \dom(\nu)$;
\item If $t$ is $prop(edge(i),l)$ then $\Theta(t,p) = \nu(\edge(p,i), l)$ when $i \leq \len(p)$, and $(\edge(i),l) \in \dom(\nu)$;
\item If $t$ is $length()$ then $\Theta(t,p) = \len(p)$.
\end{itemize}
In any other case, the evaluation of $\Theta(t,p)$ returns $\nulo$.

Let $\alpha = (t_1, \dots, t_n)$ be a tuple of projection terms and $p$ be a path. The projection of $\alpha$ over $p$, denoted $\proy{\alpha}{p}$, is the tuple
$(\Theta(t_1,p), \dots, \Theta(t_n,p))$.

\subsection{Path Algebra Operators}
Let $S$ and $S'$ be multisets of paths, $c$ be a selection condition, $\tau$ be a path restrictor, $j\geq 0$ be a natural number, and $\alpha$ be a tuple of projection terms. We define a Path Algebra composed of the following operators:
\begin{itemize}
\item \textbf{Selection:}  
$\sigma_c(S) = \{ p \in S \mid \Psi(c,p) = \true \}$
\item \textbf{Join:}  
$S \bowtie^\tau S' = \{ p_1 \circ p_2 \mid p_1 \in S \land p_2 \in S' \land p_1 \risingdotseq p_2 \land p_1 \circ p_2 \models \tau \}$
\item \textbf{Union:}  
$S \cup S' = \{ p \mid p \in S \vee p \in S' \}$
\item \textbf{Recursive Join:}  
$\phi^\tau(S) = \bigcup_{i \geq 0} \{ p \mid p \in \phi^\tau_i(S) \}$,  
where $\phi^\tau_0(S) = \sigma_\tau(S)$ and  
$\phi^\tau_i(S) = \{ p \mid p \in \phi^\tau_{i-1}(S) \lor p \in (\phi^\tau_{i-1}(S) \bowtie^\tau \phi^\tau_0(S)) \}$  
until $|\phi^\tau_i(S)| = |\phi^\tau_{i-1}(S)|$
\item \textbf{Projection:}  
$\pi^j_\alpha(S)\!=\!\{\proy{\alpha}{p_i}\!\mid\!1\!\leq\!i\!\leq\!\min(j, |S|) \}$  
where $p_i$ is the $i$-th path in $S$.
\end{itemize}

The \emph{Selection} operator $\sigma_c(S)$ filters the multiset of paths $S$ by evaluating the selection condition $c$.  
The \emph{Join} operator $S \bowtie^\tau S'$ allows the concatenation of linkable paths while enforcing the path restriction $\tau$. 
The \emph{Union} operator $S \cup S'$ corresponds to the bag union of paths, that is, the union of two multisets of paths is obtained by taking all paths from both multisets and adding their multiplicities.

The \emph{Recursive Join} operator $\phi^\tau(S)$ applies a series of join operations over the multiset of paths $S$ until the fixed-point condition $|\phi^\tau_{i}| = |\phi^\tau_{i-1}|$ is reached. Consequently, the recursive join operator $\phi^\tau$ inherits the same restriction mechanism as the join operator. Note that if the input graph contains cycles and $\tau$ is \textit{Walk}, the recursive join operator may not stop. This problem can be avoided by using any other path restrictor (\textit{Trail}, \textit{Acyclic}, or \textit{Simple}). 

Finally, the \emph{Projection} operator $\pi^j_{\alpha}(S)$ transforms a multiset of paths $S$ into a relation (i.e., a table), where each column corresponds to a projection term in $\alpha$ (maintaining the sequence order), and each row contains the projected values $\proy{\alpha}{p}$ for each path $p \in S$. If $j = 0$, all paths are projected; otherwise, only the first $j$ paths are returned.

\subsection{Path Algebra Expressions}
A path algebra expression is a syntactic construct formed by combining functions to retrieve paths from the input graph with path algebra operators. It allows us to describe complex path patterns that can be evaluated over a property graph.

A path algebra expression is defined recursively:
\begin{enumerate}
\item Given a property graph $G$, the functions $\paths_0(G)$ and $\paths_1(G)$ are simple path algebra expressions. 
\item
Let $E_1$ and $E_2$ be path algebra expressions, $\tau$ be a path restrictor, $c$ be a selection condition, $j$ be a natural number (including zero), and $\alpha$ be a sequence of projection terms.
The expressions
$(E_1 \cup E_2)$,
$(E_1 \bowtie^{\tau} E_2)$,
$\phi^{\tau}( E_1)$,
$\sigma_c( E_1)$, and
$\pi^j_{\alpha}( E_1)$
are complex path algebra expressions.
\end{enumerate} 

The evaluation of a path algebra expression $E$ over a graph $G = (\mathcal{N}, \mathcal{E}, \rho, \lambda, \nu)$, denoted $\eval(E, G)$, returns a multiset of paths defined inductively as follows:
\begin{itemize}
\item 
$\eval(\paths_0(G), G) = \{ [n] \mid n \in N \}$ where $[n]$ denotes the path obtained from a node $n$; 
\item 
$\eval(\paths_1(G), G) = \{ [n_1,e,n_2] \mid e \in E \land \rho(e) = (n_1,n_2) \}$ where $[n_1,e,n_2]$ denotes the path obtained from an edge $e$.
\item 
$\eval(\sigma_c(E_1), G) = \sigma_c( \eval(E_1, G))$
\item 
$\eval( E_1 \cup E_2, G) = \eval(E_1, G) \cup \eval(E_2, G)$
\item 
$\eval( E_1 \bowtie^{\tau} E_2, G) = \eval( E_1, G) \bowtie^{\tau} \eval( E_2, G)$
\item 
$\eval( \phi^{\tau}(E_1), G) = \phi^{\tau}(\eval(E_1, G))$
\item 
$\eval(\pi^j_{\alpha}(E_1), G) = \pi^j_{\alpha}(\eval(E_1, G))$
\end{itemize}

Note that, unlike the other operators, the projection operator $\pi^j_{\alpha}(E)$ does not return a multiset of paths, but rather a set of values extracted from those paths. For this reason, projection should only be applied at the final stage of an expression, once the desired paths have been retrieved, constructed, and filtered.

Let us consider the following path algebra expression:
\[
\pi^{10}_{nProp(1,\mathsf{name})} \left( \sigma_{eLabel(1) = \mathsf{"knows"}} \left( \paths_1(G) \right) \right).
\]
The expression $\paths_1(G)$ returns the paths of length one.
The selection operator $\sigma_{eLabel(1) = \mathtt{knows}}$ filters the paths to those where the label of the first edge is equal to ``$\mathtt{knows}$''.
Finally, the projection operator $\pi^{10}_{nProp(1,\mathsf{name})}$ returns the property ``$\mathtt{name}$'' of the first node, for each solution path. 
The final output is restricted to the first 10 paths. 

\section{Declarative Query Language}
\label{sec:language}
This section introduces the query language used in PathDB to express regular path queries declaratively. Specifically, we describe the syntax of the language and the method for translating declarative path queries into path algebra expressions. 
Hence, the evaluation of a declarative query will be given by the evaluation of its corresponding path algebra expression. 

\subsection{Path query syntax}
\label{sec:qlov}
The syntax of the language is based on the GQL standard~\cite{gql2024}.
A Path Query is an expression of the form \code{MATCH} $\dots$ \code{WHERE} \dots \code{RETURN}.
The \code{MATCH} clause defines the path pattern, the \code{WHERE} clause specifies selection conditions, and the \code{RETURN} clause determines the query output. A query may also include modifiers such as \code{LIMIT} to restrict the number of results.
The specific syntax of the language is defined by the grammar included in the Appendix. 

Next, we give an overview of the query language syntax using the path query shown in Figure \ref{fig:query1}.   

\begin{figure}[t!]
\centering
\small
\begin{verbatim}
MATCH TRAIL 
  p = (x)-[((likes.hasCreator)+)]->(y) 
WHERE x.name = "Moe" 
RETURN y.name  LIMIT 3
\end{verbatim}
\caption{Example of a PathDB declarative path query.}
\label{fig:query1}
\end{figure}

The \code{MATCH} clause includes a path restrictor (\code{TRAIL}), a path variable (\code{p}), and a path expression.
The path expression is composed of a source node variable (\code{x}), a regular expression (\code{(likes.hasCreator)+}), and a target node variable (\code{y}). 
The language supports the following types of regular expressions: single label (e.g. \code{likes}), label negation (e.g. \code{!likes}), concatenation (e.g. \code{likes.hasCreator}), alternation (e.g. \code{likes|knows}), transitive closure (e.g. \code{knows+}), Kleene star (e.g. \code{knows*}), and optionality (e.g. \code{knows?}). 

The evaluation of a path expression produces a multiset of paths whose labels satisfy the regular expression. In our example, one or more repetitions of the sequence of labels \code{likes.hasCreator}. The restrictor \code{TRAIL} ensures that the resulting paths conform to the intended semantics, in this case, paths with no repeated edges.

The clause \code{WHERE} allows us to introduce conditions over the elements of the resulting paths. For example, \code{x.name = "Moe"} filters the solution to those paths where the source node \code{x} has the property \code{name} equal to \code{"Moe"}. The WHERE clause is optional in a query.

Finally, the \code{RETURN} clause specifies the elements to be projected, transforming the multiset of paths into a relational table. Note that the terms listed in the \code{RETURN} clause form the table header, and each row corresponds to the elements extracted from a particular path. A returned element could be an object identifier, a label, a value, a boolean value ($\true$ or $\false$), or $\nulo$. In our example, \code{y.name} returns the value of the property \code{name} for the target node \code{y}. The clause \code{LIMIT} further restricts the output to at most three rows. 

In summary, the example query retrieves the names of people who can be reached from \code{Moe} by following paths whose labels contain one or more repetitions of the concatenated edge labels \code{likes.hasCreator}.
The final result of our example query is shown in Table \ref{tab:result1}.

\begin{table}[t!]
\centering
\begin{tabular}{|c|}
\hline
\textbf{\code{y.name}} \\ \hline
\code{"Lisa"} \\ \hline
\code{"Apu"} \\ \hline
\code{"Moe"} \\ \hline
\end{tabular}
\caption{Table returned by the evaluation of the path query shown in Figure \ref{fig:query1}.}
\label{tab:result1}
\end{table}

\subsection{Translating path queries into algebra expressions}
\label{sec:semantics}
The method for translating a declarative path query into a path algebra expression is given by the composition of three path translation functions, one for each query clause (i.e. \code{MATCH}, \code{WHERE}, and \code{RETURN}).
For the sake of space, we do not present a detailed description of the translation functions but do give an example of translation.   

Let $\tau$ be a path restrictor, $R$ be a regular expression, $\omega$ be a WHERE condition, 
$\varpi$ be a tuple of return terms, and $k$ be a limit parameter.
Consider that a path query $Q$ has the following structure:
\begin{verse}
\code{MATCH} $\tau$ \code{p = (x)-[$R$]->(y)} \\
\code{WHERE} $\omega$ \code{RETURN} $\varpi$ \code{LIMIT} $k$. 
\end{verse}

Given a property graph $G$, the function $\tq(Q,G)$ translates the path query $Q$ into the path algebra expression  
$\tr(\varpi, k, \tw(\omega, \tm(R,\tau,G)))$ where:

\begin{itemize}
\item 
the function $\tm(R,\tau,G)$ converts $R$ and $\tau$ into a path algebra expression $E_M$.
\item 
the function $\tw(\omega,E_M)$ converts the WHERE condition $\omega$ into a selection condition $c$ and returns a path algebra expression $E_W$ of the form $\sigma_{c}(E_M)$.
\item 
the function $\tr(\varpi,k,E_W)$ 
converts the tuple of return terms $\varpi$ into a sequence of projection terms 
$\alpha$, and returns a path algebra expression $E_Q$ of the form $\pi^{k}_{\alpha}(E_W)$.
\end{itemize}

We next illustrate this translation process with an example.

Consider that $Q$ is the path query shown in Figure \ref{fig:query1} and $G$ is the property graph shown in Figure~\ref{fig:graph}.
We have that $\tau$ is \textit{Trail}, $R$ is the regular expression \code{(likes.hasCreator)+}, $\omega$ is the where condition \code{x.name = "Moe"}, $\varpi$ is the return term \code{y.name}, and the limit parameter $k$ is equal to $3$.  

First, we apply the translation function $\tm(R,\tau,G)$ to obtain the path algebra expression
\[
E_M : \phi^{Trail}((\sigma_{label(edge(1)) = \texttt{likes})}(\paths_1(G)) \bowtie^{Trail}
\]
\[
(\sigma_{label(edge(1)) = \texttt{hasCreator})}(\paths_1(G)).
\]
Second, we use the translation function $\tw(\omega,E_M)$ to obtain the path algebra expression 
\[ 
E_W : \sigma_{prop(node(1),\texttt{name}) = \texttt{"Moe"}}(E_M).
\]
where the selection condition $prop(node(1),\texttt{name}) = \texttt{"Moe"}$ is obtained by translating the WHERE condition $\code{x.name="Moe"}$.

Finally, we apply the translation function $\tr(\varpi, k, E_W)$ to obtain the path algebra expression
\[
E_Q : \pi^{3}_{prop(node(len()+1),\texttt{name})}(E_W),
\]
where the projection term $prop(node(len()+1),\texttt{name})$ is obtained by translating the return term $\code{y.name}$.

The final path algebra expression is presented in Figure \ref{fig:qtree1} using a query tree representation.
As in relational databases, a query tree is composed of Leaves, Internal Nodes, and the Root node. Leaves are functions that return paths from the database; in our case, $\paths_0(G)$ and $\paths_1(G)$. The internal nodes represent path algebra operations (e.g., selection, join, recursive join, union, and projection).
The root node represents the final result of the query.
The paths flow from the leaves up to the root. Each internal node takes the set of paths from its children, performs an operation, and passes the resulting set of paths upwards.

\begin{figure}[t!]
\centering
\begin{forest}
    [$\pi_{prop(node(len()+1),\text{name})}^{3}$
        [$\sigma_{prop(node(1),\texttt{name}) = \texttt{"Moe"}}$
            [$\phi^{Trail}$
                [$\bowtie^{Trail}$
                    [$\sigma_{label(edge(1)) = \texttt{likes}}$
                        [$\paths_1(G)$]
                    ]
                    [$\sigma_{label(edge(1)) = \texttt{hasCreator}}$
                        [$\paths_1(G)$]
                    ]
                ]
            ]
        ]
    ]
\end{forest}
\caption{Query tree that represents the path algebra expression of the declarative path query shown in Figure \ref{fig:query1}.}
  \label{fig:qtree1}
\end{figure}

\section{Experimental evaluation}
\label{sec:expeval}
The objective of our experimental evaluation is to analyze the effectiveness of the algebra-based strategy used by PathDB. In particular, our aim is to assess whether representing query execution as compositions of algebraic operators over multisets of paths provides practical advantages over traditional traversal-driven evaluation strategies. In this sense, we compare PathDB with traversal-based baseline algorithms to isolate the impact of the proposed path algebra on query evaluation.  

\subsection{PathDB-CLI}
PathDB is currently distributed as a JAR file and provides a Command Line Interface (CLI) for data loading and query execution.
The source code, the release, and the documentation are available in the PathDB GitHub repository (\url{https://github.com/dbgutalca/PathDB}).

After downloading the current PathDB release, the user can run \code{java -jar PathDB.jar} to start the system with the default configuration.
In this case, the system automatically loads the property graph shown in Figure \ref{fig:graph} into the main memory. Then, the user can introduce and execute a path query directly.

\begin{figure} []
\centering
\includegraphics[width=8cm]{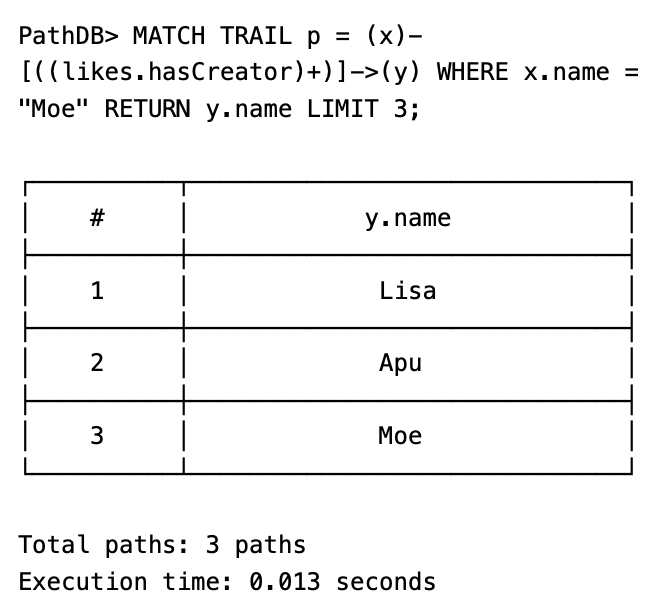}
\caption{Result of executing a path query in the  Command Line Interface (CLI) of PathDB.}
\label{fig:exec1}
\end{figure}

Figure \ref{fig:exec1} shows the result of the execution of our running query in PathDB-CLI. The query result is displayed in the console in tabular format.
In addition, the interface reports the number of solution paths and the query execution time.

PathDB-CLI supports data loading to create a custom graph.
In this case, the user must provide two files: one with the nodes' information and another with the edges' information. These files must follow the syntax defined by the Property Graph Data Format (PGDF)~\cite{10734106}.

\subsection{Baseline algorithms}
The baseline algorithm for evaluating regular path queries is automata-based graph traversal, often called the product-graph algorithm~\cite{Mendelzon1995}.
The algorithm works as follows: the query regular expression is first translated into a finite automaton; the input graph and the automaton are then traversed simultaneously by exploring the product graph whose states are pairs of graph nodes and automaton states; an edge is followed whenever the graph label matches an automaton transition, and a query result (i.e., a path) is produced when a product state with an accepting automaton state is reached.

We implemented two variants of the automata-based approach: DFS+A and BFS+A, which differ in the search strategy used to explore the product graph. DFS+A performs a depth-first traversal, while BFS+A uses breadth-first search.
Both baselines were designed to support the same class of regular path queries as PathDB and enforce identical path restrictors (Walk, Trail, Simple, and Acyclic). No indexing or pruning techniques beyond those strictly required by the restrictors were applied, ensuring that the comparison focuses exclusively on the evaluation strategy rather than optimizations.

Note that DFS+A and BFS+A represent standard traversal-based techniques for evaluating regular path expressions and therefore provide a suitable baseline for comparison with the algebra-based evaluation strategy implemented in PathDB.

\subsection{Experimental Setup}
This section describes the experimental environment and methodology used to evaluate PathDB. In particular, we detail the dataset, query workload, evaluation metrics, and hardware specifications. These elements were selected to enable a controlled evaluation of the proposed algebra-based evaluation strategy and to ensure a fair comparison with traversal-based approaches.

\subsubsection{Datasets}
As evaluation datasets, we used four property graphs produced with SNB Datagen, the data generator included in the LDBC Social Network Benchmark (SNB) \cite{91677}, a widely used benchmark to evaluate graph database systems. The SNB Datagen produces synthetic yet realistic social network graphs characterized by heterogeneous node and edge types, skewed degree distributions, and a rich set of structural patterns \cite{erling2015,szarnyas2022}. 
These characteristics make it particularly suitable for evaluating recursive and path-oriented queries, which are central to the evaluation of regular path queries.

The SNB Datagen provides a parameter that allows one to generate social network graphs with different scale factors. 
We generated four graphs with scale factors 0.1, 0.3, 1.0, and 3.0 respectively.
Basic information about these graphs is presented in Table \ref{tab:graphs}.
All experiments were conducted with the entire graphs resident in memory, ensuring that query execution times reflect computational costs rather than disk I/O.

\begin{table}[t!]
\centering
\begin{tabular}{|c|c|c|c|c|c|}
\hline
\textbf{Graph} & \textbf{Scale Factor} & \textbf{\#Nodes} & \textbf{\#Edges} & \textbf{Disk Size} \\ 
\hline
G1 & 0.1 & 327,588 & 1,477,965 & 115MB \\
G2 & 0.3 & 908,224 & 4,583,118 & 254MB \\
G3 & 1.0 & 3,181,724 & 17,256,038 & 1.3GB \\ 
G4 & 3.0 & 9,281,922 & 52,695,734 & 2.1GB \\
\hline
\end{tabular}
\caption{Basic information about the property graphs used in the experimental evaluation.}
\label{tab:graphs}
\end{table}

The SNB Datagen uses a property graph schema in which Person nodes are connected by \code{knows} relationships and interact via \code{Posts} and \code{Comments}. Content is organized in \code{Forums} and annotated with \code{Tags}, while additional nodes, such as \code{Places} and \code{Organizations}, provide context. Nodes and edges include properties and timestamps to support realistic queries. The property graph shown in Figure \ref{fig:graph} gives an idea of the evaluation dataset. 


A graph produced with SNB Datagen includes both static and dynamic graph elements. Static elements remain unchanged across scale factors, while dynamic elements grow as the graph scales. In our experiments, we primarily focused on edge labels occurring in the dynamic portion of the graph, as these labels induce higher path multiplicities and deeper recursion, allowing us to better observe the performance impact of complex path patterns and recursive algebra operators.

Additionally, SNB Datagen produces graphs containing a substantial number of cycles and multi-hop relationships, which are essential for evaluating regular path queries involving recursion, such as transitive closure and Kleene star operators. This characteristic is particularly important for assessing the behavior of different path restrictors (Walk, Trail, Simple, and Acyclic) and the scalability of recursive evaluation strategies.


\subsubsection{Query Workload}
The query workload used in our experimental evaluation was designed to systematically exercise different constructs of regular path queries. In particular, the workload focuses on patterns derived from the fundamental operators of regular expressions, such as concatenation, disjunction, recursion, and optionality, which are commonly used when expressing recursive path queries.

To structure the workload generation process, we distinguish three levels of query abstraction: Abstract Queries, Template Queries, and Real Queries.

An Abstract Query is essentially a pattern that represents a specific operation within regular expressions, such as concatenation, disjunction, recursion, or optionality. For example, the expression $(A \cdot B)*$ denotes the concatenation of $A$ and $B$, followed by the application of the Kleene star. It is important to note that $A$ and $B$ represent variables that will later be replaced by valid labels of the test graph.  

A Template Query is obtained by replacing the variables in an abstract query with valid edge labels from the graph schema. For example, if we consider the labels \code{likes} and \code{hasCreator}, then the abstract query $(A \cdot B)*$ can generate the template query \code{(likes.hasCreator)*}.

Finally, a Real Query is a concrete query that can be executed in a graph database system's native query language. Following our example, a real query in the PathDB query language will be: \code{MATCH} \code{TRAIL} \code{p = (x)-[(likes.hasCreator)*]-(y)} \code{WHERE} \code{x.id = "p3378"} \code{RETURN} \code{p}. Note that this query filters the solution to paths where the source node's identifier is \code{p3378}.

Table \ref{tab:workload} summarizes the abstract queries used in the workload together with the number of real queries generated for each pattern.
The idea was to create a diverse set of abstract queries by combining all the operators supported in a regular expression, resulting in 26 abstract queries.
For each abstract query, we defined 2-6 template queries based on the edge labels of the test graph. Each template query was then instantiated into a real query by selecting a source node whose out-degree, with respect to the corresponding adjacent label, was located at the median of the distribution.  
As we can see, the query workload yielded 142 real queries.

\begin{table}[t]
\centering
\small
\begin{tabular}{|c|c|c|}
\hline
\begin{tabular}[c]{@{}c@{}}Abstract\\ query\end{tabular} & \begin{tabular}[c]{@{}c@{}}{\#}Real\\queries\end{tabular} & Template query example \\ \hline
$A?$ & 6  & hasCreator? \\ 
$A+$ & 3 & knows+ \\ 
$A*$ & 3 & knows* \\ 
$(A \cdot B)$ & 6 & (hasModerator . knows) \\
$(A \mid B)$ & 6 & (likes | knows) \\ 
$(B \mid A)$ & 6 & (workAt | likes) \\
$(A \cdot B)?$ & 6 & (knows . likes)? \\ 
$(A \cdot B)+$ & 2 & (likes . hasCreator)+ \\ 
$(A \cdot B)*$ & 2 & (likes . hasCreator)* \\ 
$(A \mid B)?$ & 6 & (workAt | hasInterest)? \\ 
$(A \mid B)+$ & 6 & (studyAt | isLocatedIn)+ \\
$(A \mid B)*$ & 6 & (workAt | knows)* \\ 
$(A? \cdot B)$ & 6 & (hasCreator? . isLocatedIn) \\
$(A+ \cdot B)$ & 6 & (replyOf+ . hasTag) \\
$(A* \cdot B)$ & 6 & (knows* . likes) \\
$(A \cdot B?)$ & 6 & (hasMember . knows?) \\ 
$(A \cdot B+)$ & 6 & (replyOf . replyOf+) \\ 
$(A \cdot B*)$ & 6 & (hasMember . knows*) \\ 
$(A? \mid B)$ & 6 & (knows? | studyAt) \\
$(A+ \mid B)$ & 6 & (replyOf+ | hasCreator) \\
$(A* \mid B)$ & 6 & (replyOf* | hasTag)\\ 
$(A \mid B?)$ & 6 & (isLocatedIn | hasInterest?)\\
$(A \mid B+)$ & 6 & (likes | replyOf+)\\
$(A \mid B*)$ & 6 & (likes | replyOf*)\\
$((A \cdot B) \cdot C)$ & 6 & ((hasModerator . knows) . isLocatedIn)\\
$((A \mid B) \mid C)$ & 6 & ((likes | knows) | hasInterest)\\ \hline
\hline
\textbf{Total}        & 142  &                 \\ \hline
\end{tabular}
\caption{Summary of the abstract queries that comprise the query workload used in the experimental evaluation.}
\label{tab:workload}
\end{table}

\subsubsection{Evaluation Metrics}
For each real query, we measured the query execution time, defined as the elapsed time between query submission and the production of the final result set. Each query was executed three times, and the reported execution time is the average across these runs.

To ensure comparability across systems and avoid excessive memory consumption, query results were limited to at most 100 paths. A 120-second timeout was applied to each query execution. These limits were consistently enforced across all experiments.

\subsubsection{Experimental environment.}
The experiments were conducted on two virtual machines (VM) running on Proxmox VE with the following configuration:
\begin{itemize}
\item CPU: x86-64 virtual CPU (QEMU/KVM), x86-64-v2 with AES
\item Memory: 16 GB DDR4 (VM1) and 66 GB (VM2)
\item Storage: SSD
\item Operating System: Ubuntu 22.04
\item Java Virtual Machine: OpenJDK 21.0.9
\end{itemize}
PathDB and the baseline algorithms were evaluated on the same hardware platform to ensure fair and consistent measurements.

\subsection{Experimental results}


\subsubsection{PathDB versus baselines}
The first experiment implies a performance evaluation of DFS+A, BFS+A and PathDB.
For this, the real queries of the workload were executed on each system, using the graphs G1, G2, and G3, and running over the 16GB RAM virtual machine.
The average execution times of this experiment are shown in Table \ref{tab:exp1}.

It is important to note that this experiment was limited to queries using the \textit{Trail} restrictor. 
Recall that using a restrictor (except Walk) requires a special verification step, which increases the query execution time. However, the cost of this verification does not affect the overall execution time as much as the query evaluation strategy (traversal-based versus algebra-based).

Among traversal strategies, DFS+A outperforms BFS+A, achieving lower execution times on average (2.988 s versus 3.908 s for G1, 10.930 s versus 16.827 s for G2, and 31.874 s versus 36.038 s for G3). This suggests that BFS+A incurs greater overhead due to the need to maintain recursion control and path verification structures, which makes it less efficient. Consequently, DFS+A consistently delivered faster results, while BFS+A lagged behind despite its breadth-first exploration strategy.

Overall, PathDB consistently outperforms both traversal-based baselines across all queries, often by a substantial margin. 
For the minimum execution times, PathDB shows 0.173 s, BFS+A shows 0.401 s, and DFS+A shows 0.433 s.
For the highest times, we can observe 6.959 s for PathDB, 48.939 s for DFS+A, and 88.112 s for BFS+A.

The observed performance gap can be attributed to fundamental differences in evaluation strategies. Traversal-based algorithms explore the product graph formed by the input graph and the query automaton. During this exploration, the algorithms repeatedly revisit overlapping subpaths and states, resulting in redundant computations and increased execution time. In contrast, PathDB evaluates queries using algebraic operators that explicitly represent path composition and recursion. This formulation enables the systematic reuse of intermediate results, reducing redundant exploration, and improving overall efficiency.

The experimental results confirm that the proposed path algebra provides a feasible and efficient alternative to traditional traversal-based evaluation strategies for regular path queries, particularly in the presence of recursion and restrictive path semantics.
By expressing recursive path queries as compositions of algebraic operators, PathDB avoids much of the redundant exploration inherent to traversal-based approaches.

\begin{table}[!]
\caption{Average execution times (in seconds) of DFS+A, BFS+A, and PathDB for abstract queries. The real queries were evaluated on three property graphs (G1, G2, G3) of increasing scale factor (0.1, 0.3 and 1.0).
\textit{T/O} denotes a timeout exceeding 120 seconds.
The table also includes the minimum, median, average and maximum execution times.
}
  \label{tab:exp1}
  \centering
  \scriptsize
  \setlength{\tabcolsep}{5pt}
  \begin{tabular}{|c|ccc|ccc|ccc|}
    \hline

    &
    \multicolumn{3}{c|}{\textbf{G1 (SF 0.1)}} &
    \multicolumn{3}{c|}{\textbf{G2 (SF 0.3)}} &
    \multicolumn{3}{c|}{\textbf{G3 (SF 1.0)}} \\

    \cline{2-4} \cline{5-7} \cline{8-10}

    \textbf{Abstract Query} &
    DFS+A & BFS+A & PathDB &
    DFS+A & BFS+A & PathDB &
    DFS+A & BFS+A & PathDB \\

    \hline

    $A?$
      & 0.747 & 0.704 & 0.212
      & 1.292 & 1.310 & 0.550
      & 14.804 & 15.376 & 1.916 \\

    $A+$
      & 1.901 & 1.787 & 0.264
      & 6.713 & 6.539 & 0.760
      & 25.612 & 26.101 & 3.191 \\

    $A*$
      & 1.972 & 1.772 & 0.302
      & 6.567 & 6.501 & 0.841
      & 24.072 & 25.281 & 3.724 \\

    $(A \cdot B)$
      & 2.243 & 2.237 & 0.293
      & 8.534 & 7.910 & 0.804
      & 29.288 & 31.733 & 2.949 \\

    $(A \mid B)$
      & 1.468 & 1.396 & 0.335
      & 7.161 & 7.029 & 0.985
      & 29.143 & 30.720 & 2.839 \\

    $(B \mid A)$
      & 1.109 & 0.995 & 0.336
      & 5.079 & 5.034 & 0.950
      & 24.646 & 25.798 & 3.527 \\

    $(A \cdot B)^{?}$
      & 2.204 & 2.179 & 0.277
      & 7.710 & 8.079 & 0.825
      & 28.871 & 32.718 & 2.993 \\

    $(A \cdot B)+$
      & 10.282 & 4.927 & 0.419
      & 36.637 & 69.014 & 1.432
      & \textit{T/O} & 68.641 & 4.152 \\

    $(A \cdot B)*$
      & 10.094 & 4.755 & 0.460
      & 36.448 & 69.046 & 1.538
      & \textit{T/O} & 70.365 & 4.552 \\

    $(A \mid B)?$
      & 1.539 & 1.406 & 0.426
      & 7.315 & 7.227 & 1.069
      & 29.249 & 31.185 & 2.973 \\

    $(A \mid B)+$
      & 2.170 & 2.041 & 0.624
      & 4.109 & 4.033 & 1.941
      & 41.179 & 43.236 & 6.492 \\

    $(A \mid B)*$
      & 2.112 & 1.980 & 0.645
      & 8.305 & 8.292 & 1.931
      & 40.855 & 42.962 & 6.959 \\

    $(A? \cdot B)$
      & 2.177 & 2.017 & 0.278
      & 7.747 & 8.006 & 1.059
      & 29.864 & 32.214 & 3.073 \\

    $(A+ \cdot B)$
      & 5.655 & 23.025 & 0.375
      & 19.597 & 46.836 & 1.139
      & 38.797 & 88.112 & 3.988 \\

    $(A* \cdot B)$
      & 5.650 & 22.591 & 0.376
      & 19.732 & 50.873 & 1.169
      & 36.679 & 40.596 & 4.082 \\

    $(A \cdot B?)$
      & 2.166 & 1.945 & 0.293
      & 7.343 & 7.277 & 0.881
      & 27.987 & 29.510 & 2.975 \\

    $(A \cdot B+)$
      & 3.591 & 3.355 & 0.442
      & 12.417 & 12.683 & 1.374
      & 43.118 & 47.067 & 4.301 \\

    $(A \cdot B*)$
      & 3.525 & 3.285 & 0.460
      & 9.675 & 9.587 & 1.279
      & 46.689 & 48.133 & 4.447 \\

    $(A? \mid B)$
      & 1.462 & 1.344 & 0.376
      & 6.727 & 6.661 & 0.922
      & 25.667 & 27.631 & 3.501 \\

    $(A+ \mid B)$
      & 3.683 & 3.269 & 0.339
      & 12.588 & 12.387 & 0.939
      & 47.100 & 48.915 & 3.782 \\

    $(A* \mid B)$
      & 1.829 & 1.215 & 0.173
      & 6.460 & 4.449 & 0.521
      & 25.202 & 18.042 & 2.018 \\

    $(A \mid B?)$
      & 0.433 & 0.401 & 0.362
      & 6.746 & 6.661 & 0.912
      & 25.777 & 26.851 & 3.387 \\

    $(A \mid B+)$
      & 1.974 & 1.820 & 0.429
      & 6.546 & 6.427 & 1.358
      & 24.955 & 25.826 & 4.130 \\

    $(A \mid B*)$
      & 2.040 & 1.890 & 0.447
      & 9.288 & 9.523 & 1.453
      & 28.526 & 26.892 & 4.276 \\

    $((A \cdot B) \cdot C)$
      & 3.924 & 7.751 & 0.372
      & 15.772 & 48.518 & 1.064
      & 48.939 &  3.918 & 4.027 \\

    $((A \mid B) \mid C)$
      & 1.731 & 1.519 & 0.512
      & 7.678 & 7.591 & 1.243
      & 27.968 & 29.175 & 5.130 \\

    \hline
    \hline
$min$ & 0.433 & 0.401 & 0.173
    & 1.292 & 1.310 & 0.521
    & 14.804 & 3.918 & 1.916 \\
$med$ & 2.139 & 1.963 & 0.374
    & 7.694 & 7.751 & 1.062
    & 29.007 & 30.953 & 3.753 \\
$avg$ & 2.988 & 3.908 & 0.378
    & 10.930 & 16.827 & 1.13
    & 31.874 & 36.038 & 3.822 \\
$max$ & 10.282 & 23.025 & 0.645
    & 36.637 & 69.637 & 1.941
    & 48.939 & 88.112 & 6.959 \\    

    \hline    

  \end{tabular}
\end{table}

\subsubsection{PathDB under different path restrictors}
The second experiment aims to evaluate the performance of PathDB under different path restrictors, that is, \textit{Walk}, \textit{Trail}, \textit{Simple}, and \textit{Acyclic}.
For this, the workload test queries were executed in PathDB on the G4 test graph (the biggest), and consequently we used the 64GB RAM virtual machine.
The average execution times of this experiment are shown in Table \ref{tab:exp2}, and a graphical comparison can be observed in Figure \ref{fig:exp2}.

The results show that PathDB maintains consistent performance across most abstract query patterns. On average, non-recursive regular expressions are generally executed in less than one second, while more complex abstract query patterns involving repetition or closure operators exceed three seconds. This indicates that the system scales reasonably well, although the syntactic complexity of regular expressions directly impacts execution time.

When comparing the different path restrictors, we observe that:   
\textit{Simple} shows the lowest minimum time with 0.639 seconds for $(A \mid B)$;
\textit{Walk} shows the lowest median and average times with 2.474 and 6.488 seconds respectively;
\textit{Acyclic} shows the longest time with 36.511 seconds for $(A \cdot B)+$;

The most resource-intensive queries are abstract query patterns involving recursive operators, such as
$(A \cdot B)*$ for \textit{Walk}, and $(A \cdot B+)$ for the other three path restrictors. 
In contrast, union-based query patterns such as $(A \mid B)$ and $(B \mid A)$ are executed very efficiently, with times below 0.8 seconds.

In general terms, the behavior of execution times is consistent with theoretical expectations based on the complexity of abstract query patterns. This confirms that complexity is a relevant variable in query execution time. For example, the abstract query pattern $(A \cdot B)$ has an execution time of 1.436 seconds for \textit{Walk}, whereas adding an additional concatenation increases the time to 1.269 seconds. 
A considerably greater increase is observed when recursion is introduced: the pattern $(A \cdot B)*$ reaches a time of 28.384 seconds, which represents a significant difference from the non-recursive pattern $(A \cdot B)$.

Another relevant aspect is the internal implementation of path operations in PathDB. Since PathDB represents each path as an object that contains all node and edge references, memory consumption increases with the complexity of the operation. In the case of a simple concatenation such as $(A \cdot B)$, it is only necessary to store the multisets of paths $A$ and $B$ in memory and then perform the join operation. However, in abstract query patterns that contain recursion, the calculation becomes more complex, since not only must the multisets of paths of $A$ and $B$ be maintained, but also the intermediate results generated by the recursive expansions, which implies greater memory requirements and, consequently, longer execution times.

During the evaluation, errors associated with memory consumption were also observed.
PathDB showed significantly fewer errors than the baselines. Specifically, these errors occur with abstract queries of the form $(A \cdot B)+$ and $(A \cdot B)*$, where the set of intermediate paths becomes extremely large. These errors can be resolved by allocating additional system memory or by optimizing the implementation of the recursive operator to reduce memory consumption.

Our analysis suggests that optimizing operators -- particularly the recursive operator -- could significantly reduce execution times and improve overall system efficiency. Since abstract query patterns involving recursion incur the highest computational costs, a more efficient implementation of this type of operator would directly impact PathDB's performance.

In summary, PathDB works well in simple queries but is sensitive to the complexity of regular expressions and the chosen path restrictor. For applications with intermediate-complexity queries, PathDB offers competitive, consistent performance. In scenarios involving complex regular expressions, the choice of path restrictor and the limitation of recursions becomes critical: \textit{Walk} and \textit{Trail} are efficient, while \textit{Simple} and \textit{Acyclic} tend to be more difficult.

\begin{table}[t!]
  \caption{Execution Times (in seconds) of PathDB under four path restrictors (Walk, Trail, Simple and Acyclic). The test queries were evaluated on the property graph G4 (scale factor 3.0) and using the 64~GB RAM virtual machine. The table also includes the minimum, median, average and maximum execution times for each  path restrictor.}
  \label{tab:exp2}
  \centering
  \small
  \begin{tabular}{|c|cccc|}
    \hline

    & \multicolumn{4}{c|}{\textbf{G4 (SF 3.0)}} \\

    \cline{2-5}

    \textbf{Abstract Query} & Walk & Trail & Simple & Acyclic \\

    \hline

    $A?$
      & 1.540 & 1.259 & 1.232 & 1.433 \\

    $A+$
      & 5.909 & 7.252 & 5.711 & 5.010 \\

    $A*$
      & 9.106 & 6.486 & 7.576 & 5.836 \\

    $(A \cdot B)$
      & 1.436 & 1.149 & 1.237 & 1.303 \\

    $(A \mid B)$
      & 0.880 & 0.913 & 0.681 & 0.882 \\

    $(B \mid A)$
      & 0.765 & 0.896 & 0.619 & 0.889 \\

    $(A \cdot B)?$
      & 2.267 & 2.521 & 1.869 & 2.135 \\

    $(A \cdot B)+$
      & 26.453 & 22.603 & 20.598 & 20.200 \\

    $(A \cdot B)*$
      & 28.384 & 25.577 & 27.749 & 21.239 \\

    $(A \mid B)?$
      & 1.386 & 1.201 & 1.132 & 1.227 \\

    $(A \mid B)+$
      & 2.326 & 2.090 & 0.512 & 0.639 \\

    $(A \mid B)*$
      & 3.331 & 2.966 & 3.555 & 3.822 \\

    $(A? \cdot B)$
      & 2.103 & 2.301 & 1.660 & 1.888 \\

    $(A+ \cdot B)$
      & 1.437 & 3.009 & 3.183 & 3.309 \\

    $(A* \cdot B)$
      & 4.846 & 5.383 & 3.961 & 3.988 \\

    $(A \cdot B?)$
      & 2.622 & 2.640 & 1.942 & 2.311 \\

    $(A \cdot B+)$
      & 8.265 & 28.337 & 35.189 & 36.551 \\

    $(A \cdot B*)$
      & 19.476 & 24.917 & 29.264 & 28.589 \\

    $(A? \mid B)$
      & 1.816 & 1.565 & 1.559 & 1.718 \\

    $(A+ \mid B)$
      & 6.711 & 6.524 & 5.372 & 6.746 \\

    $(A* \mid B)$
      & 6.847 & 4.746 & 3.180 & 5.501 \\

    $(A \mid B?)$
      & 1.705 & 1.517 & 1.518 & 1.667 \\

    $(A \mid B+)$
      & 7.252 & 5.659 & 5.712 & 5.751 \\

    $(A \mid B*)$
      & 18.463 & 9.079 & 11.637 & 13.713 \\

    $((A \cdot B) \cdot C)$
      & 2.104 & 1.935 & 1.865 & 1.999 \\

    $((A \mid B) \mid C)$
      & 1.269 & 1.295 & 0.954 & 1.277 \\

    \hline
    \hline    
$min$ & 0.765 & 0.896 & 0.512 & 0.639 \\
$med$ & 2.474 & 2.803 & 2.561 & 2.810 \\
$avg$ & 6.488 & 6.685 & 6.903 & 6.909 \\
$max$ & 28.384 & 28.337 & 35.189 & 36.551 \\

    \hline    
    
  \end{tabular}
\end{table}

\begin{figure*}[!]
  \centering
  \begin{tikzpicture}
    \begin{axis}[
      ybar,
      bar width        = 3pt,
      width            = 1\textwidth,
      height           = 7.2cm,
      enlarge x limits = 0.025,
      ybar=0.5pt, 
      xlabel           = {Query Pattern},
      ylabel           = {Execution Time (s)},
      xlabel style     = {font=\small, yshift=-4pt},
      ylabel style     = {font=\small},
      ymin = 0,
      ymax = 40,
      ytick = {0,5,10,15,20,25,30,35,40},
      yticklabel style = {font=\scriptsize},
      xtick            = {1,2,3,4,5,6,7,8,9,10,11,12,13,
                          14,15,16,17,18,19,20,21,22,23,24,25,26},
      xticklabels = {
        $A^{?}$,
        $A^{+}$,
        $A^{*}$,
        $(A{\cdot}B)$,
        $(A{|}B)$,
        $(B{|}A)$,
        $(A{\cdot}B)^{?}$,
        $(A{\cdot}B)^{+}$,
        $(A{\cdot}B)^{*}$,
        $(A{|}B)^{?}$,
        $(A{|}B)^{+}$,
        $(A{|}B)^{*}$,
        $(A^{?}{\cdot}B)$,
        $(A^{+}{\cdot}B)$,
        $(A^{*}{\cdot}B)$,
        $(A{\cdot}B^{?})$,
        $(A{\cdot}B^{+})$,
        $(A{\cdot}B^{*})$,
        $(A^{?}{|}B)$,
        $(A^{+}{|}B)$,
        $(A^{*}{|}B)$,
        $(A{|}B^{?})$,
        $(A{|}B^{+})$,
        $(A{|}B^{*})$,
        $({(A{\cdot}B)}{\cdot}C)$,
        $({(A{|}B)}{|}C)$
      },
      xticklabel style = {
        rotate    = 70,
        anchor    = east,
        font      = \scriptsize,
        inner sep = 1.0pt,
      },
      ymajorgrids      = true,
      grid style       = {dashed, gray!35},
      axis on top      = false,
      tick pos         = left,
      legend style = {
        at             = {(0.5,-0.42)},
        anchor         = north,
        legend columns = 4,
        column sep     = 8pt,
        font           = \small,
        draw           = black,
        fill           = white,
        /tikz/every even column/.append style = {column sep=4pt},
      },
      legend cell align = left,
    ]

    \addplot[
      fill  = clrWalk,
      draw  = clrWalk!80!black,
      fill opacity = 0.85,
    ] coordinates {
      (1,1.540)(2,5.909)(3,9.106)(4,1.436)(5,0.880)(6,0.765)
      (7,2.267)(8,26.453)(9,28.384)(10,1.386)(11,2.326)(12,3.331)
      (13,2.103)(14,1.437)(15,4.846)(16,2.622)(17,8.265)(18,19.476)
      (19,1.816)(20,6.711)(21,6.847)(22,1.705)(23,7.252)(24,18.463)
      (25,2.104)(26,1.269)
    };
    \addlegendentry{Walk}

    \addplot[
      fill  = clrTrail,
      draw  = clrTrail!80!black,
      fill opacity = 0.85,
    ] coordinates {
      (1,1.259)(2,7.252)(3,6.486)(4,1.149)(5,0.913)(6,0.896)
      (7,2.521)(8,22.603)(9,25.577)(10,1.201)(11,2.090)(12,2.966)
      (13,2.301)(14,3.009)(15,5.383)(16,2.640)(17,28.337)(18,24.917)
      (19,1.565)(20,6.524)(21,4.746)(22,1.517)(23,5.659)(24,9.079)
      (25,1.935)(26,1.295)
    };
    \addlegendentry{Trail}

    \addplot[
      fill  = clrSimple,
      draw  = clrSimple!80!black,
      fill opacity = 0.85,
    ] coordinates {
      (1,1.232)(2,5.711)(3,7.576)(4,1.237)(5,0.681)(6,0.619)
      (7,1.869)(8,20.598)(9,27.749)(10,1.132)(11,0.512)(12,3.555)
      (13,1.660)(14,3.183)(15,3.961)(16,1.942)(17,35.189)(18,29.264)
      (19,1.559)(20,5.372)(21,3.180)(22,1.518)(23,5.712)(24,11.637)
      (25,1.865)(26,0.954)
    };
    \addlegendentry{Simple}

    \addplot[
      fill  = clrAcyclic,
      draw  = clrAcyclic!80!black,
      fill opacity = 0.85,
    ] coordinates {
      (1,1.433)(2,5.010)(3,5.836)(4,1.303)(5,0.882)(6,0.889)
      (7,2.135)(8,20.200)(9,21.239)(10,1.227)(11,0.639)(12,3.822)
      (13,1.888)(14,3.309)(15,3.988)(16,2.311)(17,36.551)(18,28.589)
      (19,1.718)(20,6.746)(21,5.501)(22,1.667)(23,5.751)(24,13.713)
      (25,1.999)(26,1.277)
    };
    \addlegendentry{Acyclic}

    \end{axis}
  \end{tikzpicture}

  \caption{Visual comparison of the execution times obtained by PathDB, for different query patterns, and using four path restrictors (Walk, Trail, Simple, and Acyclic). The specific times are presented in Table \ref{tab:exp2}}
  \label{fig:exp2}

\end{figure*}

\subsection{Threats to Validity}
\label{sec:threats}

\textbf{Internal validity.}
We implemented the traversal baselines (DFS+A and BFS+A) in this work.
Although we validated the implementations on small graphs and verified that they enforce the same path restrictors as PathDB, implementation choices (e.g., data structures, caching, and object allocation) may still influence performance.
To mitigate this threat, we used the same programming language and runtime, applied identical restrictor checks, and avoided indexing or pruning beyond what is required by each restrictor.

\smallskip
\noindent
\textbf{Construct validity.}
Our primary metric is the end-to-end query execution time.
This metric captures practical performance, but may conflate distinct costs, including automaton construction, path restrictor verification, intermediate-result materialization, and garbage collection.
We mitigate this threat by averaging results across multiple queries per pattern and by evaluating a workload that systematically exercises regular-expression operators (concatenation, disjunction, optionality, and recursion).

\smallskip
\noindent
\textbf{External validity.}
We used four LDBC SNB property graphs of different scale factors using an in-memory engine.
Although SNB provides realistic structure and abundant cycles, our results may not be generalized to other graph topologies, label distributions, or disk-backed engines.
In future work, we will expand the evaluation to additional scale factors, datasets, and storage settings.

\smallskip
\noindent
\textbf{Validity of the conclusion.}
We report aggregate results over a finite workload (142 instantiated queries derived from 26 abstract patterns), which may not cover all real-world query distributions.
Different label choices or source-node selections may lead to different path multiplicities and recursion depths.
To reduce sensitivity, we selected source nodes using a median out-degree heuristic and included multiple template instantiations for each abstract pattern.

\section{Related Work}
\label{sec:relatedwork}
Path queries are a core operation in graph database systems. These queries aim to determine whether there exists a path between two nodes that satisfies specific conditions.
In this section, we focus on RPQ evaluation and algebraic methods for path querying.
RPQs form the basis of property-path features in widely used graph query languages and standards, where a regular expression constrains the sequence of edge labels along a path.
Consequently, a central challenge is balancing expressive path semantics with efficient evaluation and manageable result sizes.


\subsection{Regular Path Query Evaluation}
Typically, evaluating a regular path query returns pairs of nodes connected by a path whose label sequence matches the regular expression. Some applications, however, require enumerating the matching paths, which motivates distinguishing methods that return node pairs from those that return full paths. Because the number of matching paths can be exponential in the size of the graph, full-path evaluation techniques typically rely on additional constraints (e.g., path restrictors, shortest-path variants, or bounds) to keep enumeration tractable.

Techniques that return node pairs include automata products \cite{Mendelzon1995}, automata-guided DFS/BFS \cite{Bonifati2018}, path indexing \cite{Fletcher2016}, distributed (grid-aware) evaluation \cite{Miao2007}, fine-grained complexity analysis \cite{Casel2023}, compressed matrix operations \cite{10.1007/s00778-024-00885-6}, streaming-graph evaluation \cite{10.1145/3318464.3389733}, and \textmu-RA algebra \cite{10.1145/3318464.3380567}.
Most of these approaches operate under walk semantics; however, \cite{Mendelzon1995} uses simple-path semantics (node uniqueness), while \cite{10.1145/3318464.3389733} supports walks, trails, simple paths, and shortest paths.
Regardless of semantics, these techniques primarily target reachability and often do not reconstruct the actual paths. They typically rely on automata products, recursive joins, matrix-based methods, or pruning techniques to evaluate RPQs efficiently.

In contrast, methods that return full paths include PathFinder \cite{10.1007/978-3-031-77850-6_8}, its GQL adaptation \cite{gql2024}, and G-CORE \cite{angles2018}. PathFinder and GQL offer fine-grained control over path semantics and can enumerate multiple paths, whereas G-CORE returns a single shortest walk that satisfies the query pattern. These approaches enable full path reconstruction, making them suitable for applications that require explicit traversal information.
Table \ref{tab:rpq-minimal-summary} summarizes key RPQ evaluation approaches, highlighting supported path semantics and the nature of the returned results. The table also highlights that supporting multiple restrictors while returning full paths is less common than endpoint-pair evaluation under walk semantics.

\begin{table}[h!]
\centering
\footnotesize
\caption{Summary of RPQ Evaluation Techniques by Path Restriction and Output}
\begin{tabular}{|c|c|c|c|c|c|c|}
\hline
\textbf{Ref.} & \textbf{Walk} & \textbf{Trail} & \textbf{Simple} & \textbf{Acyclic} & \textbf{Returns} \\
\hline
\cite{Mendelzon1995} &  &  & \checkmark &  & Node pairs \\
\hline
\cite{Bonifati2018} & \checkmark &  &  &  & Node pairs \\
\hline
\cite{Fletcher2016} & \checkmark &  &  &  & Node pairs \\
\hline
\cite{Miao2007} & \checkmark &  &  &  & Node pairs \\
\hline
\cite{Casel2023} & \checkmark &  &  &  & Node pairs \\
\hline
\cite{10.1007/s00778-024-00885-6} & \checkmark &  &  &  & Node pairs \\
\hline
\cite{10.1145/3318464.3389733} & \checkmark & \checkmark & \checkmark &  & Node pairs \\
\hline
\cite{10.1145/3318464.3380567} & \checkmark &  &  &  & Node pairs \\
\hline
\cite{angles2018} & \checkmark &  &  & & Full paths \\
\hline
\cite{gql2024} & \checkmark & \checkmark & \checkmark & \checkmark & Full paths \\
\hline
\cite{10.1007/978-3-031-77850-6_8} & \checkmark & \checkmark & \checkmark & \checkmark & Full paths \\
\hline
PathDB & \checkmark & \checkmark & \checkmark & \checkmark & Full paths \\
\hline
\end{tabular}
\label{tab:rpq-minimal-summary}
\end{table}

\subsection{Algebras for Path Querying}
Several studies have explored algebraic structures for expressing and computing path operations. Gondran \cite{Gondran1975} proposes an adjacency-matrix algebra with addition (union of paths sharing the same endpoints) and multiplication (path concatenation). This framework supports reachability via Kleene closure, path counting, and optimization tasks (e.g., shortest paths, k-best paths, and maximum capacity) in static graphs. However, it returns only reachable node pairs (not full paths) and assumes arbitrary path semantics.

Manger \cite{Manger2004} introduces a path algebra with two operators: union (grouping paths that converge at a node) and product (concatenating paths with matching endpoints). This approach simplifies the syntax and supports explicit path reconstruction, but it remains limited to arbitrary semantics and may struggle in cyclic graphs.

Rodriguez and Neubauer \cite{90388} present a path algebra with four operators: union (grouping paths with identical endpoints), join (concatenation via a shared intermediate node), Cartesian product (combining unrelated paths), and Kleene closure (generating arbitrary-length paths via automata). Their model supports complex path construction and manipulation.

Additionally, graph databases such as Neo4j \cite{Neo4j2020} provide basic path operations: create (build a path from a start node and relationships), combine (concatenate paths with matching endpoints), split (extract subpaths), and decompose (return ordered sequences of nodes and relationships).
In contrast to prior path algebras that often rely on arbitrary semantics and provide limited support for explicit path reconstruction, our approach introduces a formally grounded algebra for precise and expressive path manipulation.
Unlike earlier formulations that treat paths mainly as witnesses for reachability, our algebra is closed over multisets of paths, making intermediate path collections first-class values that can be composed and reused across operators.
Gondran’s model focuses on reachability and optimization without reconstructing paths, whereas Manger's algebra enables reconstruction but struggles with cycles. Our algebra supports explicit path generation while handling cycles robustly. Compared to Rodriguez and Neubauer's framework, which emphasizes combinatorial expressiveness, our algebra aims to balance expressiveness with formal semantics. Finally, unlike Neo4j's operational primitives, our model provides a declarative and extensible algebraic foundation suitable for both theoretical analysis and practical implementation in dynamic graph environments.

\section{Conclusions and Future Work}
\label{sec:conclusion}
In this paper, we present PathDB, a modular in-memory query engine for evaluating Regular Path Queries (RPQs) over property graphs. The core contribution of PathDB is a closed path algebra over multisets of paths, providing a formal, compositional framework for expressing recursive path computations. PathDB translates declarative path queries into algebraic execution plans composed of iterator-based operators, enabling systematic reuse of intermediate results and flexible enforcement of path semantics.

The experimental evaluation shows that PathDB consistently outperforms automaton-guided DFS and BFS traversal baselines across all 26 abstract query patterns, confirming that the algebraic formulation reduces redundant computation inherent to traversal-based strategies. Additionally, we demonstrate that PathDB is capable of answering queries using different path restrictors (Walk, Trail, Simple, and Acyclic). 

The current implementation has several limitations that point to directions for future work. The recursive operator is the primary performance bottleneck, particularly for patterns involving Kleene Star and transitive closure over concatenated expressions, where intermediate path sets can grow very large. Optimizing this operator, for example, by using lazy evaluation or result pruning, will improve performance. Additionally, the current physical plan manager uses a simple one-to-one mapping from logical to physical operators; introducing a cost-based optimizer with operator reordering and predicate pushdown could yield significant performance gains. Future work also includes extending the query language to support multi-path queries and aggregation over paths, scaling PathDB to disk-backed storage for graphs that exceed available RAM, and conducting a more comprehensive evaluation across multiple scale factors and graph topologies.

\section*{Data and Code Availability}
\label{sec:availability}

\textbf{Code.}
The PathDB source code, releases, and documentation are available at: \url{https://github.com/dbgutalca/PathDB}.
To support reproducibility, we also provide the baseline executables and the query workload: \url{https://github.com/dbgutalca/pathdb-eval1}.

\smallskip
\noindent
\textbf{Data.}
The evaluation dataset is generated using the LDBC Social Network Benchmark (SNB) data generator.
Due to the size of the generated graph, the links to the dataset files are available at: \url{https://github.com/dbgutalca/pathdb-eval1}



\paragraph*{Declaration of generative AI and AI-assisted technologies in the writing process}
During the preparation of this article, the authors used \emph{ChatGPT} (GPT-5.3) and \emph{Claude} (Sonnet 4.6) to edit and improve grammar. After using these tools, the authors reviewed and edited the content as needed and assume full responsibility for the content of the publication.

\bibliographystyle{IEEEtran}
\bibliography{biblio}

\appendix

\section{PathDB Declarative Query Language Grammar.}
\label{sec:grammar}

\begin{Verbatim}[fontsize=\small]
<pathQuery>   ::= <matchClause> <whereClause> <returnClause>
<matchClause> ::= "MATCH" <restrictor>? <pathPattern>    
<restrictor>  ::= "WALK" | "TRAIL" | "ACYCLIC' | "SIMPLE"
<pathPattern> ::= <var> "=" <pathExp> 
<pathExp>     ::= "("<sVar>?")-["<regEx>"]("{"..number"}")?->("<tVar>?")"
<regEx>       ::= <label> | !<label> | <regEx> "." <regEx> | <regEx> "|" <regEx> | 
                  <regEx>"+"| <regEx>"*"| <regEx>"?"
<whereClause> ::= ("WHERE" <condition>)?
<condition>   ::= <term> <comparisonOp> <value> | <boolTerm> "("<condition>")" | 
                  <condition> <logicalOp> <condition>  
<returnClause>::= "RETURN" <rterm> (,<rterm>)*  <limit>?
<rterm>       ::= <boolTerm>|<term>                   
<limit>       ::= "LIMIT" number
<term>        ::= <object> | "LABEL("<object>")"| <object>"."<prop>| "LENGTH()"
<object>      ::= "NODE("<index>")" | "EDGE("<index>")" | "FIRST()" | "LAST()" | <sVar> | 
                  <tVar>
<boolTerm>    ::= "ISTRAIL()" | "ISSIMPLE()" | "ISACYCLIC()"
<comparisonOp>::= "=" | "!=" | "<" | ">" | "<=" | ">="  
<logicalOp>   ::= "AND" | "OR" 
<index>       ::= [1-9] [0-9]*
\end{Verbatim}

\end{document}